# An FFT approach to the analysis of dynamic properties of gas/liquid interfaces

Sandrine Mariot[a,*], Valentin Leroy[b], Juliette Pierre[b], Florence Elias[b,c], Eloïse Bouthemy[a], Dominique Langevin[a], Wiebke Drenckhan[a]

[a] *Laboratoire de Physique des Solides, Université Paris-Sud – UMR8502, France*
[b] *Matière et Systèmes Complexes, Université Paris 7 – UMR7057, France*
[c] *UPMC Université Paris 06, UFR 925, France*

The characterisation of the dynamic properties of viscoelastic monolayers of surfactants at the gas/liquid interface is very important in the analysis and prediction of foam stability. With most of the relevant dynamic processes being rapid (thermal fluctuation, film coalescence etc.) it is important to probe inter- facial dynamics at high deformation rates. Today, only few techniques allow this, one of them being the characterisation of the propagation of electro-capillary waves on the liquid surface. Traditionally, this technique has been applied in a continuous mode (i.e. at constant frequency) in order to ensure reliable accuracy. Here we explore the possibility to analyse the propagation of an excited pulse in order to access the interfacial properties in one single Fourier treatment over a wide range of frequencies. The main advantage of this approach is that the measurement times and the required liquid volumes can be reduced significantly. This occurs at the cost of precision in the measurement, due partly to the presence of a pronounced resonance of the liquid surface. The pulsed approach may therefore be used to pre-scan the surface response before a more in-depth scan at constant frequency; or to follow the changes of the interfacial properties during surfactant adsorption.

1. Introduction

Efforts of several generations of researchers to elucidate the stability of liquid foams have brought out clearly the importance of the dynamic properties of the gas/liquid interfaces. These contain monolayers of interfacially active foaming agents, which may be low molecular weight surfactants, proteins, polymers or even particles. The presence of these agents tunes the interfacial tension and, in many cases, renders the interfaces viscoelastic. This viscoelastic response is controlled by the relaxation dynamics within the interface and/or between the interface and the bulk, rendering the response to an interfacial deformation strongly frequency-dependent. A multitude of techniques has therefore been developed in the past in order to access different types of interfacial deformation (dilation and/or shear) at different ranges of frequencies. While in the low frequency range, reliable techniques are now available (oscillating bubble, interfacial shear rheometer etc.) the higher frequency range of several hundred Hz remains difficult to probe. These frequency ranges are particularly important for questions of thermal fluctuations and for the rupture of the thin foam films. One of the most commonly used techniques is that of the propagation of an excited electro- capillary wave [1], where the analysis of the wave profile away from the excitation provides access to the dynamic properties of the interface. However, this technique has the disadvantage that acquisition times are long, since the wave profile has to be scanned for each frequency individually. This makes it impossible, for example, to follow the variation of the interfacial response during surfactant adsorption. Moreover, the measurement requires large liquid surfaces in order to avoid wave reflection from the sur- face border, which renders the experiment sensitive to evaporation issues and expensive when working with costly interfacially active agents.

Here we report our investigations into the use of a pulsed electro-capillary excitation, which, via Fourier analysis, provides immediate access to a wide range of frequency. For this purpose, we excite an electrocapillary wave at one point. A laser scan of the passing wave at two well-known distances from the excitation allows us to obtain the wave velocity and dissipation as a function of frequency from a single pulse – and therefore the complete dissipation relation. The passage of a single pulse taking only a few hundred milliseconds, one can dramatically reduce the typical measurement times in comparison to the continuous technique. The time which is gained, however, also leads to a loss in precision. A compromise is achieved by averaging over many pulses, which increases measurement times to a few seconds or minutes, but which remains short in comparison to the continuous technique. Moreover, the accessible frequency ranges are reduced due to the reduced energy input in the pulsed mode, which leads to a rapid dissipation of the higher frequency parts of the spectrum. This reduction in energy is amplified by the presence of a well-defined resonance of the liquid surface in the low frequency range (order 10 Hz with the liquids investigated here).

At the outset of this article (Section 2) we present briefly the key theoretical ingredients of wave propagation at gas/liquid interfaces with and without the presence of monolayers. We then present in detail the experimental set-up and the data analysis (Section 3). In Section 4.1 we discuss in some depth the measurements on pure liquid interfaces before presenting some results obtained with surfactant monolayers (Section 4.2).

## 2. Theory

An excited, propagating capillary wave of frequency $\omega$ and wavelength $\lambda$ on the surface of a liquid can be described by a damped wave profile

$$A(x,t) = A_0 \exp[i(kx - \omega t)], \tag{1}$$

where $k = k_r + ik_i$ is the complex wave vector with $k_i$ being the damping coefficient and $k_r = 2\pi/\lambda$. The phase velocity $v$ of the wave is then given by

$$v = \frac{\omega}{k_r}. \tag{2}$$

The description of the dispersion relation for such waves has to couple the fluid dynamics of the bulk with the boundary conditions imposed the gas/liquid interface. If this interface contains a monolayer of surface active species it may have visco-elastic properties whose influence can be taken into account by considering a frequency-dependent viscoelastic modulus

$$E = E_r + iE_i. \tag{3}$$

Here $E_r$ is the real part, corresponding to the surface elasticity, and $E_i$ is the imaginary part, being related to a surface viscosity $\eta_S$ via $E_i = \omega\eta_S$.

After full hydrodynamic treatment [2, 3], the dispersion relation of the problem may be written as

$$[\gamma k^3 + \rho g k + i\eta\omega k(k+m) - \rho\omega^2][Ek^2 + i\eta\omega(k+m)] + \eta^2\omega^2 k(k-m)^2 = 0, \tag{4}$$

where $\rho$ is the density of the liquid, $\gamma$ the surface tension and $g$ the gravitational acceleration. $k$ and $m$ are related via

$$m^2 = k^2 + i\frac{\omega\rho}{\eta}. \tag{5}$$

The sign of $m$ must be taken so that the solution of the dispersion relation is physically realistic [1,2]. A close analysis of Eq. (4) [1,2] shows that in order to have sufficient precision in the measurement of the moduli, the elastic modulus of the interface needs to be of the order of 15% of the surface tension. The physical origin of this condition is a result of the fact that in order to differentiate between the different surface effects, the transverse and longitudinal wave propagation need to be in resonance. This puts an important limit on the applicability of the technique.

In the absence of a monolayer, or in the limit of negligible interfacial visco-elasticity one may set $E = 0$ in Eq. (4). Assuming furthermore that the bulk viscosity $\eta$ is small, the dispersion relation of Eq. (4) simplifies to

$$\omega^2 = \frac{\gamma k^3}{\rho} + gk + i\frac{\omega\eta k^2}{\rho}. \tag{6}$$

If dissipation is negligible, this relation reduced to the well-known Kelvin equation, which allows to relate $\omega$ and $k_r$ and therefore to describe the variation of the phase velocity $v$ with $\omega$

$$\omega^2 = \frac{\gamma k_r^3}{\rho} + g k_r. \qquad (7)$$

From this equation one can conclude that gravitational effects can be neglected when

$$k_r^{-1} \ll \sqrt{\frac{\gamma}{\rho g}} = l_c, \qquad (8)$$

i.e. when the inverse wave vector is much smaller than the capillary length *lc*. In our experiments, this is typically the case for $f = \omega/2\pi > 15\,\text{Hz}$.

In the remaining article, the data will always be plotted using the corresponding numerical solutions of Eq. (6) (i.e. $E = 0$) as reference. For the calculation of the corresponding visco-elastic moduli we use the numerical solution of the full dispersion relation given in Eq. (4).

### 3. Materials and methods

*3.1 Experimental setup*

The specificity of the electro-capillary wave technique is that the generation and the detection of the wave occur without any mechanical contact with the surface of the studied solution. Our set-up is inspired by the one designed by Stenvot and Langevin [3] and Sohl and Ketterson [4], which we improved through automa- tion and mathematical analysis tools, as described in the following.

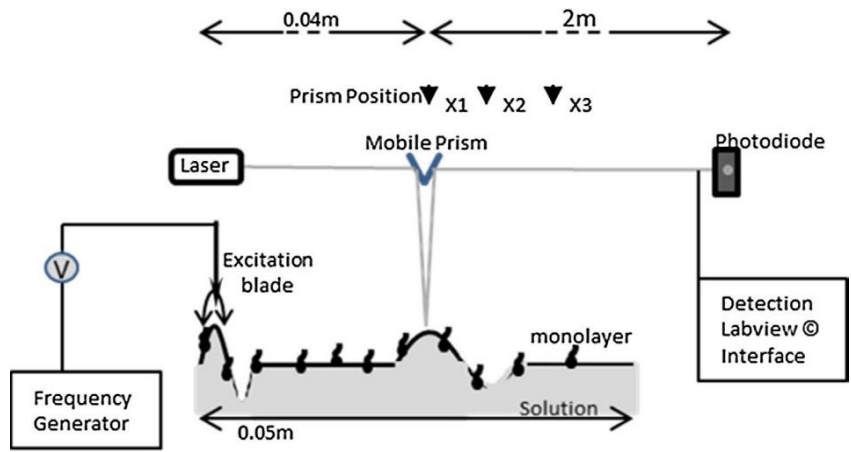

**Fig. 1.** Sketch of the experimental set-up: an electrocapillary wave is excited on the monolayer by applying a voltage of 230 V between a long razor blade and the liquid surface. The propagating wave is detected by the reflection of a laser beam.

As sketched in Fig. 1, the wave is generated on the liquid surface by electrocapillary effects [1,2]. In our case, we apply a sinusoidal voltage of 230 V and frequency $f_{ex}$ between a long razor blade (length $L_b$ = 10 cm, height $h_b$ = 1 cm, width $w_b$ = 0.02 cm) and the monolayer on the air–liquid interface. The blade is placed parallel and very close to the surface (0.1 mm). As the dielectric constant of the water is higher that the dielectric constant of air ($\varepsilon_w$ = 80 >> $\varepsilon_0$ = 1), the liquid is attracted by the blade. Since this attraction depends on the direction of the field gradient, and not of the field, the excitation of the surface wave occurs at $f_0$ = 2$f_{ex}$. The excitation is performed by a frequency generator (homemade, Labview® Controler) which allows us to choose between different wave forms.

The excitation results in the generation of planar waves which propagate away from the razor blade. As shown in Fig. 1, the detection of the wave profile is performed by a laser (Melles Griot, 5 mW, 635 nm), which runs parallel to the liquid surface and which is reflected from the surface at a chosen point by a mirror system. Before reflection off the surface, the beam is focalised by optical lenses. After reflection it reaches a position-sensitive detector (PSD). The detection device is placed far enough (~2m) in order to amplify the small deflections. The mirror system is motorised by a displacement stage and can scan the surface over a distance of 5 cm with a precision of 0.01 mm. Amplitudes of the excited waves are of the order of 1 µm, which means that in the range of wavelengths investigated here, the wave amplitude is directly proportional to the deflection of the laser beam on the position sensitive detector.

In our experiments we excite short pulses in a periodic mode. Their propagation is scanned at two fixed positions $X_1$ and $X_2$ (notation [$X_1$, $X_2$]). At each position we average over 50 detected pulses in order to optimise the precision. The averaged pulses of both positions are then treated using a classic Fast Fourier Transform (FFT) analysis with a home-made programme in order to obtain the dispersion relation of the wave. The treatment is described in more detail in Section 3.2.

For historic reasons the liquid is contained in a large Teflon trough with the following dimensions: length $L$ = 50 cm, width $w$ = 13 cm, depth $h$ = 0.7 cm. With the pulsed approach a much smaller trough – and hence much less solution – can be used. This is discussed in some more detail in the conclusions (Section 5). In order to avoid external vibrations which can disturb the surface, we isolated the device thanks to an anti-vibrational table. The large dimensions are chosen so that reflections from the trough boundaries can be neglected.

As is obvious from Eq. (4), a proper description of the dispersion relation requires a precise measurement of the surface tension. We perform these measurements using a DeltaPi device from Kibron, which has a sensitivity of 10 µN/m. This device is based on the Wilhelmy method [5] using a small diameter probe (0.51 mm) made of a special alloy wire (commercialised by Kibron) which is perfectly wetted by the liquid. Even though the sensitivity of the measurement is very high, the reproducibility of the Wilhelmy technique is slightly lower and indicated by the errors of the surface tension values provided in Table 1.

All experimental in-put and measurement parameters (excitation frequency and voltage, scanning positions, beam position, profile detection, etc.) are automated and managed via a Labview interface.

### 3.2 Analysis

The response-profile of the liquid surface is collected at a fixed position $X_n$ in terms of the amplitude variation of the wave as a function of time: $A_n(X_n, t)$. Fig. 2 shows an example of a propagating pulse on pure water at two positions $X_1 = 10$ mm and $X_2 = 25$ mm from the razor blade. These two profiles are then treated by a Fast Fourier Transform analysis in order to obtain the frequency-dependent amplitude $A_n(f)$ and phase $\phi(f)$ of the signal.

Fig. 2b and c shows the phase shift $\phi_2(f) - \phi_1(f)$ and the amplitude $A_n(f)$ of the signals obtained after FFT treatment. These can be directly used to obtain the frequency-dependent damping coefficient $k_i(f)$ via

$$k_i(f) = \frac{1}{X_2 - X_1} ln\left[\frac{A_1(f)}{A_2(f)}\right], \tag{9}$$

and the phase velocity by

$$v(f) = \frac{2\pi f(X_2 - X_1)}{\varphi_2(f) - \varphi_1(f)}. \tag{10}$$

All dispersion relations in the following article will be presented via $k_i$ and $v$.

### 3.3 Materials

As reference systems we use pure water (Milli-Q system) and a water/glycerol mixture which contains 35% glycerol which was mixed for 1 h before use. The glycerol (Sigma–Aldrich, pure > 99.5%) was used as provided by Sigma Aldrich.

**Table 1** Physico-chemical properties of the different samples investigated in this article.

| Solution | Concentration | Surface Tension (mN/m)(@25°c) | Viscosity (Pa*s) | Density (kg/m$^3$) |
|---|---|---|---|---|
| Pure Milli-Q Water | | 72 | 0.001 | 1000 |
| 65%Pure Water+35%Glycerol | | 68 | 0.003 | 1090 |
| DTAB (Dodecyltrimethylammonium bromide) | @cmc (15mM) | 34.6 | 0.001 | 1000 |
| DTAB | @cmc/3 | 44.0 | 0.001 | 1000 |
| DTAB | @cmc*3 | 34.4 | 0.001 | 1000 |

For the preparation of the monolayers we used DTAB (dodecyltrimethylammonium bromide) solutions of different concentrations: cmc/3, cmc, cmc*3. The cmc (Critical Micelle Concentration) of DTAB is 15 mM. The DTAB was used as received from Sigma Aldrich. The solutions were freshly prepared for every experiment with ultrapure water (Milli-Q system). They were mixed by agitation during 4 h and studied 12 h later.

The physicochemical parameters of all liquids/solutions studied are shown in Table1. The viscosity of the water/glycerol mixture was measured using a rheometer. The density of the same mixture was obtained by calculation since the densities of the individual components are known.

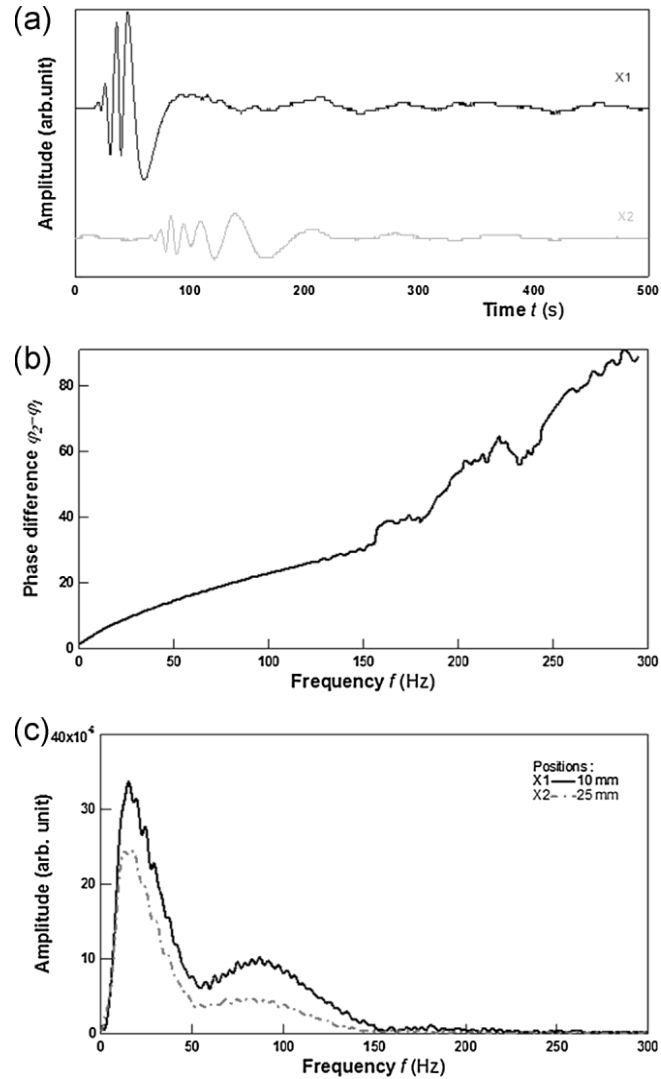

**Fig. 2.** (a) Detected wave profiles at increasing distance from the excitation ($X_1$ = 10 mm, $X_2$ = 25 mm) on pure water. (b) Phase difference as a function of frequency between both signals obtained from the FFT. (c) Amplitude spectrum obtained for both signals. (b) and (c) are used to calculate the phase velocity $v$ and the damping constant $k_i$ of the wave, respectively.

## 4. Results

*4.1. Pure liquids*

*4.1.1. Exploration of excitation and detection conditions*

In order to optimise the wave excitation, we investigated the influence of the wave profile influence of the wave profile of the excited wave on a pure water surface. We used simple sinusoidal excitations at frequency $f_{ex}$ which are cut abruptly after N periods in order to broaden their spectrum. Two examples with N = 1 and N = 5 are shown in Fig. 3a. We compare these with sinusoidal excitations which are modulated with a Gaussian envelope

$$A(t) = \exp\left(-\frac{(t-t_0)^2}{\sigma^2}\right) \cos(2\pi f_{ex}(t-t_0)) \tag{11}$$

where $f_{ex}(=f_0/2)$ is the central frequency of the signal, $t_0$ the center of the pulse and $\sigma$ the pulse width.

A representative selection of some resulting spectra (in arbitrary units) at X = 10 mm away from the excitation is shown in Fig. 3b. The first observation is that the characteristic frequency of the response is at twice the excited frequency, as discussed in Section 3.1.

The second observation is the presence of a pronounced peak at a characteristic frequency of $f_r \approx 10$ Hz. We think that this a due to a resonance behaviour of the liquid surface which may be rationalised as follows: during the excitation, the razor blade deforms the surface over a characteristic length, which is given by the balance of gravitational and capillary forces, hence by the capillary length $l_c$. If we approximate the mass of the moved volume by $l_c^3 \rho$ and say that the interface acts like a membrane of tension $\gamma$, the resonance frequency can be approximated as

$$f_r \approx \sqrt{\frac{\gamma}{l_c^3 \rho}} / (2\pi) \tag{12}$$

In the case of pure water this gives $f_r \approx 10$ Hz, while in the case of surfactant solutions this gives $f_r \approx 11$ Hz. This corresponds very well to the experimental observations. Unfortunately, these frequencies are not of particular interest here, as in this range gravity plays an important role and the wavelengths are so large that one should work with larger liquid volumes in order to avoid the influence of the bottom of the container.

Using different wave forms helps to distribute slightly differently the energies between the resonance peak and the frequency range of interest. As shown in Fig. 3b, the use of sinusoidal pulses with Gaussian envelopes allows to bring energy into a well-defined frequency range away from the resonance. However, we found that the zone of sufficient signal remained concentrated in a narrow frequency zone. The excitation with a simple sinusoidal pulse cut after one period was the most efficient mean to obtain a sufficiently energetic response over a wide frequency range. It is for this reason that in the following we use this simple pulse.

Having chosen the excitation by a simple pulse, we investigated the influence of the excitation frequency. Fig. 4a shows the response spectrum (in arbitrary units) at $X = 10$ mm away from the excitation for different excitation frequencies keeping all other conditions constant. It is apparent that the closer the excitation frequency is to the resonance frequency the more energy is transmitted into the liquid surface. Higher frequency excitations are dissipated rapidly. This dissipation is also visible when scanning the same pulse at increasing distance from the excitation (Fig. 4b).

Combining the different observations, we therefore decided to use throughout this work a simple sinusoidal pulse at 50 Hz cut after one period.

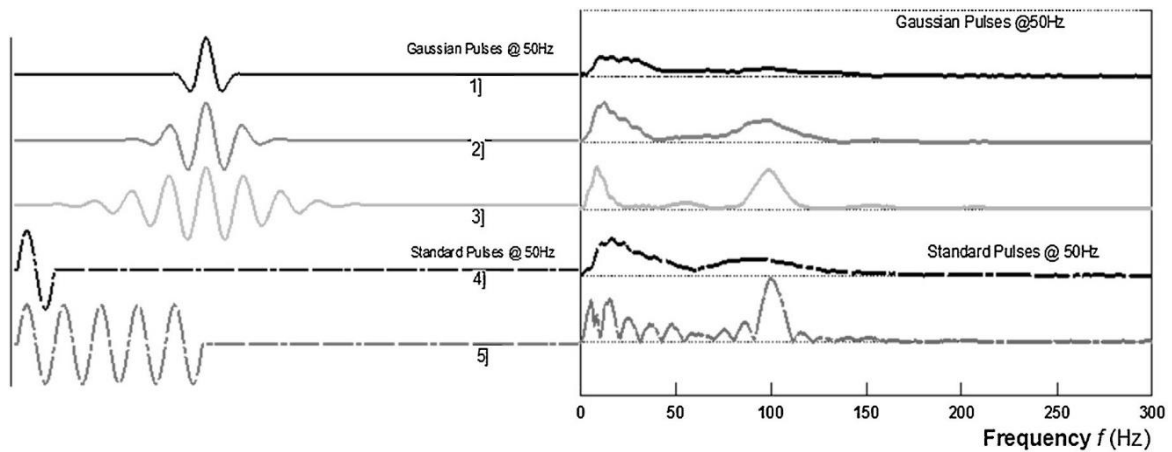

**Fig. 3.** (a) Different wave forms investigated for the excitation. (b) Resulting response of the liquid surface at 10 mm from the excitation: [1–3] Gaussian pulses with $f_{ex}$ = 50 Hz and $\sigma$ = 0.01, 0.02, 0.04 s, [4, 5] standard pulses with $f_{ex}$ = 50 Hz and $N$ = 1, 5 periods.

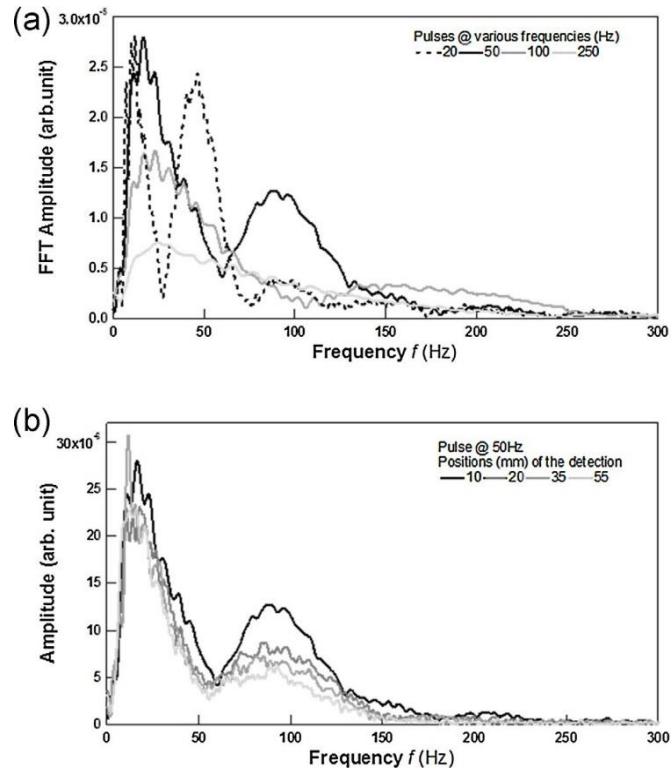

**Fig. 4.** (a) Influence of the excitation frequency on the generated wave spectrum on the surface of water and (b) change of the wave spectrum at different positions with increasing distance from the excitation.

*4.1.2. Full analysis*

In order to test the method, we investigate here the propagation of the pulse on pure liquids, in particular we compare pure water with a water/glycerol mixture in order to investigate the influence of the bulk viscosity.

Fig. 5a and b show the resulting phase velocity $v$ and the damping coefficient $k_i$, respectively, for both liquids. The lines are obtained using the real (Fig. 5a) and imaginary (Fig. 5b) part of the theoretical dispersion relation equation (Eq. (6)) using the physical parameters given in Table 1. One notices that the measurements of the phase velocity are very precise. The damping coefficients show a certain scatter. This is normal since the pure liquids dissipate only little between the two measuring points. More precision may be obtained by measuring over longer distances and/or at three, instead of two points. However, both velocity and damping agree very well with the theoretical prediction.

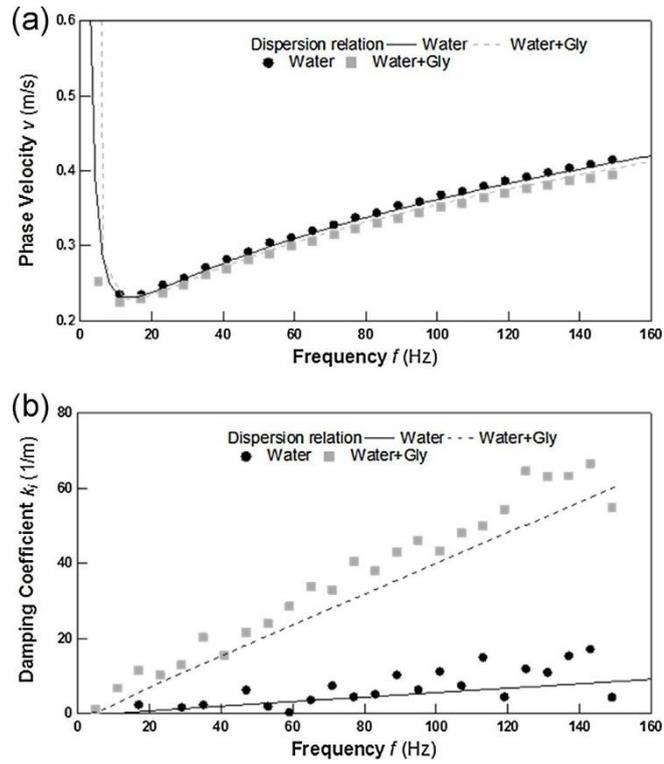

**Fig. 5.** Phase velocity $v$ (a) and damping coefficient $k_i$ (b) for water and a water/glycerol mixture. The lines are obtained from the theory given in Eq. (6) using the physical parameters given in Table 1.

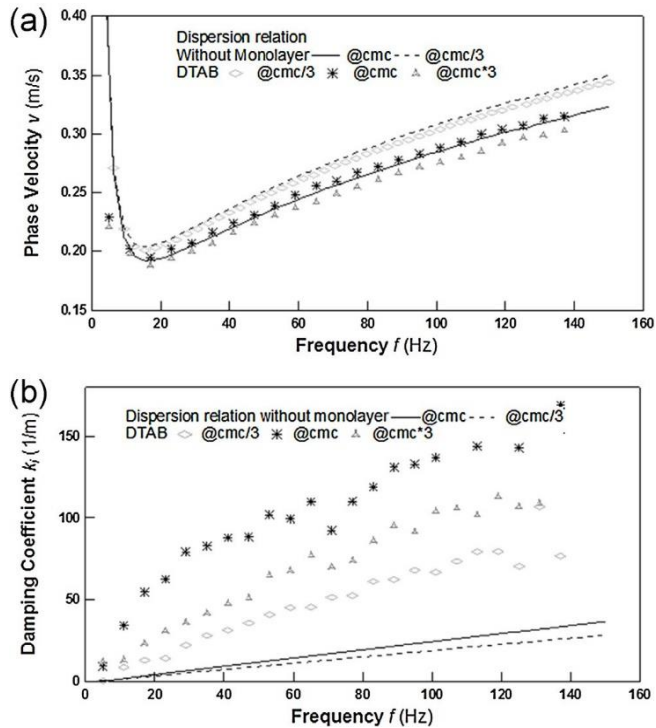

**Fig. 6.** Dispersion relation for DTAB solutions at three different concentrations (cmc, cmc/3, cmc and 3cmc). (a) Wave velocity $v$ and (b) damping coefficient as a function of frequency $f$. The lines correspond to Eq. (6), i.e. they take into account only the effect of the monolayer on the surface tension, without considering its visco-elasticity.

*4.2. Surfactant monolayers*

Here we illustrate the applicability of the pulsed capillary wave technique by analysing the wave propagation on surfactant monolayers formed by the cationic surfactant DTAB at three different surfactant concentrations: cmc/3, cmc, and cmc*3 (Section 3.3). All solutions have the same viscosity and density, while the surface tension and the visco-elastic properties of the air/liquid interface depend on the surfactant concentration. The physical parameters of the solutions are provided in Table 1.

Fig. 6 shows the phase velocity $v$ and the damping coefficient $k_i$ obtained for the three DTAB solutions. For illustrative purposes, the data is compared with Eq. (6) which describes the wave propagation in the absence of a monolayer taking into account only the influence of the monolayer on the surface tension (Table 1), but not on the visco-elastic properties. The purpose of showing them here is therefore to bring out the influence of the visco-elasticity on the wave propagation.

Fig. 6a shows that the presence of the monolayers has only a small effect on the velocity of the propagating wave, which is also confirmed by analysing Eq. (4). However, since the measurement of the wave velocity occurs with high precision, this is generally not a problem [1,2]. The damping increases strongly with the presence of the monolayers, as shown in Fig. 6b.

As one can see in Eq. (4), the presence of a non-negligible visco- elasticity of the monolayer has an influence on both the wave velocity and the damping. Moreover, elastic and viscous contributions are linked in a non-intuitive manner, since the experiment probes the coupling between longitudinal and transverse waves (Section 2). We have therefore used our data together with Eq. (4) in order to obtain a measure of the frequency-dependent visco-elastic moduli $E_r$ (elastic modulus) and $E_i$ (loss modulus) (Fig. 7).

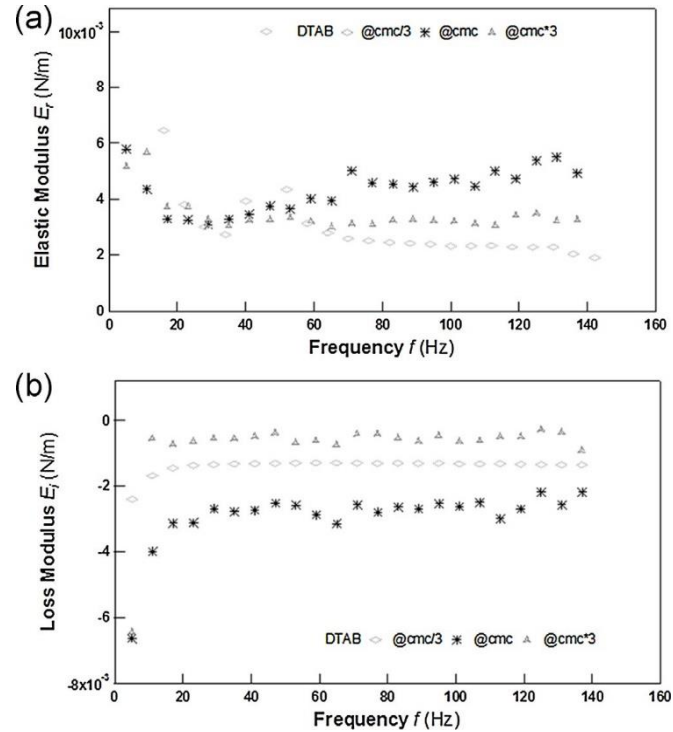

**Fig. 7.** Elastic (a) and loss (b) modulus obtained from using the data of Fig. 6 in combination with Eq. (4).

A first observation is that at low frequencies the moduli increase significantly, even though they should go to zero. We think that this may be due to different experimental side effects. One effect may come from the fact that at low frequencies the wave propagation into the liquid bulk is not damped sufficiently strongly to neglect the influence of the bottom of the trough which, in our case, is just 7 mm underneath the liquid surface. The propagation of waves on a shallow liquid needs to take into account the liquid height $h$ when the wave lengths becomes of this order. The wave length $\lambda$ being given by $v/f$, one finds that $\lambda$ is of the order of 10–40 mm in the frequency range of 5–20 Hz. The amplitude of the wave decreasing as $\exp(-2\pi z/\lambda)$, this gives values which may be non-negligible in the low frequency range. This problem could be treated by using a dispersion relation which takes into account the finite depth [6], or by simply using a deeper container. Another effect could be that due to the resonance in this frequency regime (Section 4.1.1), the wave behaviour may be non-linear. Since reliable techniques exist to investigate the low frequency behaviour [7], we did not investigate the deviations in the low-frequency range in detail. For frequencies above 20 Hz we verified that a variation of the depth of the trough and of the wave amplitude did not influence the results.

A comparison with previous work is given in Fig. 8 [7,8]. Because the frequency range of our experiments is lower, the elastic mod- ulus should be lower according to the model proposed by Levich and made popular by Lucassen [9]. However, the model is applicable for non-ionic surfactants, and it has to be implemented for ionic surfactants [10]. It is out of the scope of the present paper to apply the rigorous model to the data, it suffices to note that the modulus decreases with decreasing frequency as observed for cmc/3.

The reverse is seen for cmc and 3cmc. However, the loss modulus is negative, a feature that signals anomalous damping, frequently seen with surfactant solutions [11]. Anomalous damping was never seen for DTAB at high frequencies, but measurements with a homologous surfactant, CTAB, showed that the loss modulus was negative, with a tendency to increase (in absolute value) when the frequency decreases [12]. The anomalous values of the loss moduli could be due to a coupling between capillary and compression waves which is not properly taken into account in the dispersion relation, due to the presence of adsorption barriers or to an effective thickness of the interface which is not negligible in comparison with the small wave amplitudes. The problem is still open [11], but in any case it seems that the dispersion relation does not supply physically correct values for the viscous modulus, and the values of the elastic moduli are also most likely erroneous [11].

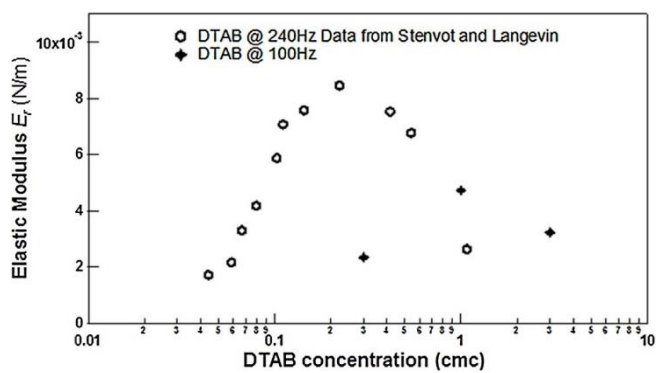

**Fig. 8.** Comparison of our results at 100 Hz with those obtained by Stenvot and Langevin [3] at 240 Hz using the continuous approach.

## 5. Conclusion

We have shown here that the propagation of pulsed electro- capillary waves on liquid surfaces is a promising route to obtain rapidly the dispersion relation over a reasonably wide frequency range. Our data fit very well with the theory on pure liquid surfaces. In the presence of monolayers we observe the expected tendencies. Reliable quantitative analysis of the dispersion relation in order to characterise the visco-elastic properties of the monolayer is sensitive to the hydrodynamic model used for the description of the problem.

Despite numerous investigations, many questions remain open in this field, especially when working with ionic surfactants [11].

Nevertheless, we think that the analysis of pulsed electro- capillary waves may provide important information about the dynamic processes at interfaces, in particular during the initial stage of surfactant adsorption at an initially bare interface. In analogy to the concept of a "dynamics surface tension", this may be thought of as a "dynamic surface visco-elasticity". Capillary waves in a continuous mode have been used successfully for this purpose in order to follow the adsorption of interfacially active polymers [6,13]. A pulsed approach would have the advantage of probing a range of frequencies simultaneously. For this purpose it would be of interest to optimise excitation and detection of the waves such that a wider frequency range may be probed.

The pulsed approach also allows to work with much smaller amounts of solutions than the continuous excitation. Since the energy input is much smaller, the waves are dissipated quite rapidly away from the excitation, so that reflection of the waves on the container walls is less of an issue. Moreover, one can use special container designs which avoid wave reflection. Such containers have typically a depth which decreases towards the wall in order to profit from the dissipation of the wave in the bulk solution.

Another promising direction may be to make use of the presence of the resonance to probe the interfacial properties. The position and width of the resonance peak may provide reliable information about the surface tension, and the elastic and dissipative contributions of the surfactants.

**Acknowledgements**

This work was funded by an ANR Grant "SAMOUSSE" (ANR- 11-BS09-001) and an ERC Starting Grant (307280-POMCAPS). The authors would like to thank Marc Rabaud and Frederic Moisy from FAST for discussions which helped to shed light on the presence of the characteristic resonance frequency. We would also like to thank the other SAMOUSSE members (Caroline Derec, Emmanuelle Rio, Benjamin Dollet, Isabelle Cantat, Arnaud Saint-Jalmes, Jérome Crassous and Cyprien Gay) for numerous discussions and valuable encouragement. We would also like to thank Frederic Restagno for his support. The authors acknowledge the help of numerous intern- ship students on this project: Marine Defossé, Sylvain Encatassamy.

**References**

[1] K. Miyano, Externally excited surface waves, in: D. Langevin (Ed.), Light Scat- tering by Liquid Surfaces and Complementary Techniques, Marcel Dekker Inc., New York, 1991.

[2] B.A. Noskov, Capillary waves in interfacial rheology, in: R. Miller, L. Liggieri (Eds.), Progress in Colloid and Interface Science, CRC Press, Boca Raton, 2009.

[3] C. Stenvot, D. Langevin, Study of viscoelasticity of soluble monolayers using analysis of propagation of excited capillary waves, Langmuir 4 (5) (1988) 1179–1183.

[4] C.H.M.K. Sohl, J.B. Ketterson, Novel technique for dynamic surface tension and viscosity measurements at liquid–gas interfaces, Rev. Sci. Instrum. 49. (10) (1978).

[5] H.-J. Butt, K. Graf, M. Kappl, Physics and Chemisty of Interfaces, Wiley, 2006.

[6] B.A. Noskov, et al., Dynamic surface properties of poly(N-isopropylacrylamide) solutions, Langmuir 20 (22) (2004) 9669–9676.

[7] R. Miller, D. Moebius, in: R. Miller, D. Moebius (Eds.), Drops and Bubbles in Interfacial Research. Studies in Interface Science, Elsevier, 1997.

[8] R. Miller, L. Liggieri, in: R. Miller, L. Liggieri (Eds.), Interfacial Rheology Progress in Colloid and Interface Science, vol. 1, CRC Press, 2009.

[9] J. Lucassen, M. Vandente, Dynamic measurements of dilational elasticity of a liquid interface, Chem. Eng. Sci. 27 (6) (1972) 1283.

[10] A. Bonfillon, D. Langevin, Electrostatic model for the viscoelasticity of ionic surfactant monolayers, Langmuir 10 (9) (1994) 2965–2971.

[11] D. Langevin, F. Monroy, Marangoni stresses and surface compression rheology of surfactant solutions. Achievements and problems, Adv. Colloid Interface Sci. 206 (0) (2014) 141–149.

[12] F. Monroy, J.G. Kahn, D. Langevin, Dilational viscoelasticity of surfactant mono- layers, Colloids Surf. A: Physicochem. Eng. Aspects 143 (2–3) (1998) 251–260.

[13] B.A. Noskov, Dynamic elasticity of triblock copolymer of poly(ethylene oxide) and poly(propylene oxide) on a water surface, Colloid J. 68 (5) (2006) 588–596.